\documentclass[showpacs,aps,prd,nofootinbib,floatfix,amsmath,amssymb]{revtex4}
\usepackage{graphicx}
\begin{document}
\makeatletter
\newbox\slashbox \setbox\slashbox=\hbox{$/$}
\newbox\Slashbox \setbox\Slashbox=\hbox{\large$/$}
\def\pFMslash#1{\setbox\@tempboxa=\hbox{$#1$}
  \@tempdima=0.5\wd\slashbox \advance\@tempdima 0.5\wd\@tempboxa
  \copy\slashbox \kern-\@tempdima \box\@tempboxa}
\def\pFMSlash#1{\setbox\@tempboxa=\hbox{$#1$}
  \@tempdima=0.5\wd\Slashbox \advance\@tempdima 0.5\wd\@tempboxa
  \copy\Slashbox \kern-\@tempdima \box\@tempboxa}
\def\FMslash{\protect\pFMslash}
\def\FMSlash{\protect\pFMSlash}
\def\miss#1{\ifmmode{/\mkern-11mu #1}\else{${/\mkern-11mu #1}$}\fi}
\makeatother

\title{Higgs mediated lepton flavor violating tau
decays $\tau \to \mu \gamma$ and $\tau \to \mu \gamma \gamma$ in
effective theories}
\author{J. I. Aranda$^{(a)}$, F. Ram\'\i rez--Zavaleta$^{(b)}$, J. J. Toscano$^{(b,c)}$, E. S.
Tututi$^{(a)}$}
\address{
$^{(a)}$Facultad de Ciencias F\'\i sico Matem\' aticas,
Universidad Michoacana de San Nicol\' as de
Hidalgo, Av. Francisco J. M\' ujica S/N, 58060, Morelia, Mich., M\' exico. \\
$^{(b)}$Facultad de Ciencias F\'{\i}sico Matem\'aticas,
Benem\'erita Universidad Aut\'onoma de Puebla, Apartado Postal
1152, Puebla, Pue., M\'exico.\\
$^{(c)}$Instituto de F\'{\i}sica y Matem\' aticas, Universidad
Michoacana de San Nicol\' as de Hidalgo, Edificio C-3, Ciudad
Universitaria, C.P. 58040, Morelia, Michoac\' an, M\' exico.}

\begin{abstract}
The size of the branching ratios for the $\tau \to \mu \gamma$ and
$\tau \to \mu \gamma \gamma$ decays induced by a lepton flavor
violating Higgs interaction $H\tau \mu$ is studied in the frame of
effective field theories. The best constraint on the $H\tau
\mu$ vertex, derived from the know measurement on the muon anomalous
magnetic moment, is used to impose the upper bounds $Br(\tau
\to \mu \gamma)<2.5\times 10^{-10}$ and $Br(\tau \to \mu \gamma
\gamma)<2.3\times 10^{-12}$, which are more stringent than current
experimental limits on this class of transitions.
\end{abstract}

\pacs{13.33.Dx, 23.20.-g, 12.60.Fr}

\maketitle

The exact conservation of lepton flavor can be considered a
central feature of the standard model (SM). However, experimental
data have shown evidence for oscillations of
atmospheric~\cite{SK}, solar~\cite{NO2}, reactor~\cite{APO}, and
accelerator neutrinos~\cite{ALI}, which suggest that lepton flavor
conservation cannot longer be taken to be granted. In a
renormalizable theory, flavor changing neutral currents can be
induced by scalar and vector fields. Although these effects can be
mediated by Higgs and $Z$ bosons at the level of classical action,
those induced by the photon field constitute a net quantum effect,
as they only can be generated at one--loop or higher orders. At
the one--loop level, electromagnetic lepton flavor violating (LFV)
transitions are induced in presence of massive
neutrinos~\cite{LFVN,MJH}. They can also be induced by LFV Higgs
couplings to leptons that naturally arise from extended Yukawa
sectors~\cite{SHER1,SHER2,DT,TT,AMM,DNR}. In this paper, we are
interested in investigating the LFV $\tau \to \mu \gamma$ and
$\tau \to \mu \gamma \gamma$ decays in the context of extended
Yukawa sectors, which are always present within the SM with
additional $SU_L(2)$--Higgs multiplets or in larger gauge groups.
Although this effect is absent in the SM, it is expected that more
complicated Higgs sectors tend to favor its presence. We will
assume that these LFV transitions of the tau lepton are mediated
by a virtual scalar field with mass of the order of the Fermi
scale $v\approx246$ GeV. However, instead of focusing in a
specific model, we will adopt a model--independent approach by
using the effective Lagrangian technique~\cite{EL}, which is an
appropriate scheme to study those processes that are suppressed or
forbidden in the SM. In this brief report, we present a
calculation for the $\tau \to \mu \gamma$ and $\tau \to \mu \gamma
\gamma$ decays induced at the one--loop level by a LFV $H\tau \mu$
vertex generated by an effective Yukawa sector that includes
$SU_L(2)\times U_Y(1)$ invariants of dimension higher than four.
It has been shown~\cite{EFYS} that a Yukawa sector extended with
dimension--six operators generates the most general coupling of
the Higgs boson to quarks and leptons without necessity of
introducing additional degrees of freedom. In particular, the
effective Yukawa sector for the leptonic sector can be written
as~\cite{DT,EFYS}
\begin{equation}
{\cal L}^Y_{eff}=-Y_{ij}(\bar{L}_i\Phi
l_j)-\frac{\alpha_{ij}}{\Lambda^2}(\Phi^\dag \Phi)(\bar{L}_i\Phi
l_j)+H.c.,
\end{equation}
where $Y_{ij}$, $L_i$, $\Phi$, and $l_i$ stand for the usual
components of the Yukawa matrix, the left--handed lepton doublet,
the Higgs doublet, and the right--handed charged lepton singlet.
The $\alpha_{ij}$ numbers are the components of a $3\times 3$
general matrix, which parametrizes the details of the underlying
physics, whereas $\Lambda$ represents the new physics scale. We
would like to emphasize that the above Lagrangian describes to the
most general $Hl_il_j$ vertex of renormalizable type, which
reproduces the main features of most extended Yukawa sectors, as
the most general version of the two Higgs doublet model
(THDM-III)~\cite{DNR} and multi--Higgs models that comprise
additional multiplets of $SU_L(2)\times U_Y(1)$ or scalar
representations of larger gauge groups. Our approach also cover
more exotic formulations of flavor violation, as the so--called
familons models~\cite{FM} or theories that involves an Abelian
flavor symmetry~\cite{AFS}. Neglecting likely CP effects, the
$H\tau \mu$ vertex can be written as $-i\Omega_{\tau \mu}$,
where~\cite{EFYS}:
\begin{equation}
\Omega_{\tau
\mu}=\frac{1}{\sqrt{2}}\Big(\frac{v}{\Lambda}\Big)^2\Big(V_L\alpha
V^\dag_R\Big)_{\tau \mu}.
\end{equation}
Here, $V_{L,R}$ are the usual unitary matrices that relate gauge
states to mass eigenstates. Our main goal in this work is to
predict model--independent upper bounds for the branching ratios
associated with the Higgs--mediated $\tau \to \mu \gamma$ and
$\tau \to \mu \gamma \gamma$ decays. As we will see below, the
best bounds are obtained using the constraint on the $\Omega_{\tau
\mu}$ parameter derived from the experimental result for the muon
anomalous magnetic moment~\cite{FPT}. It should be commented that
the one--loop contribution of this LFV vertex to the muon
anomalous magnetic moment is free of ultraviolet
divergences~\cite{FPT}, which is due to the fact that the $H\tau
\mu$ vertex has a renormalizable structure. By the same token, as
we will see below, the contribution of this vertex to the
one--loop $\tau \to \mu \gamma$ and $\tau \to \mu \gamma \gamma$
decays is free of divergences.

\begin{figure}
\centering
\includegraphics[width=3.0in]{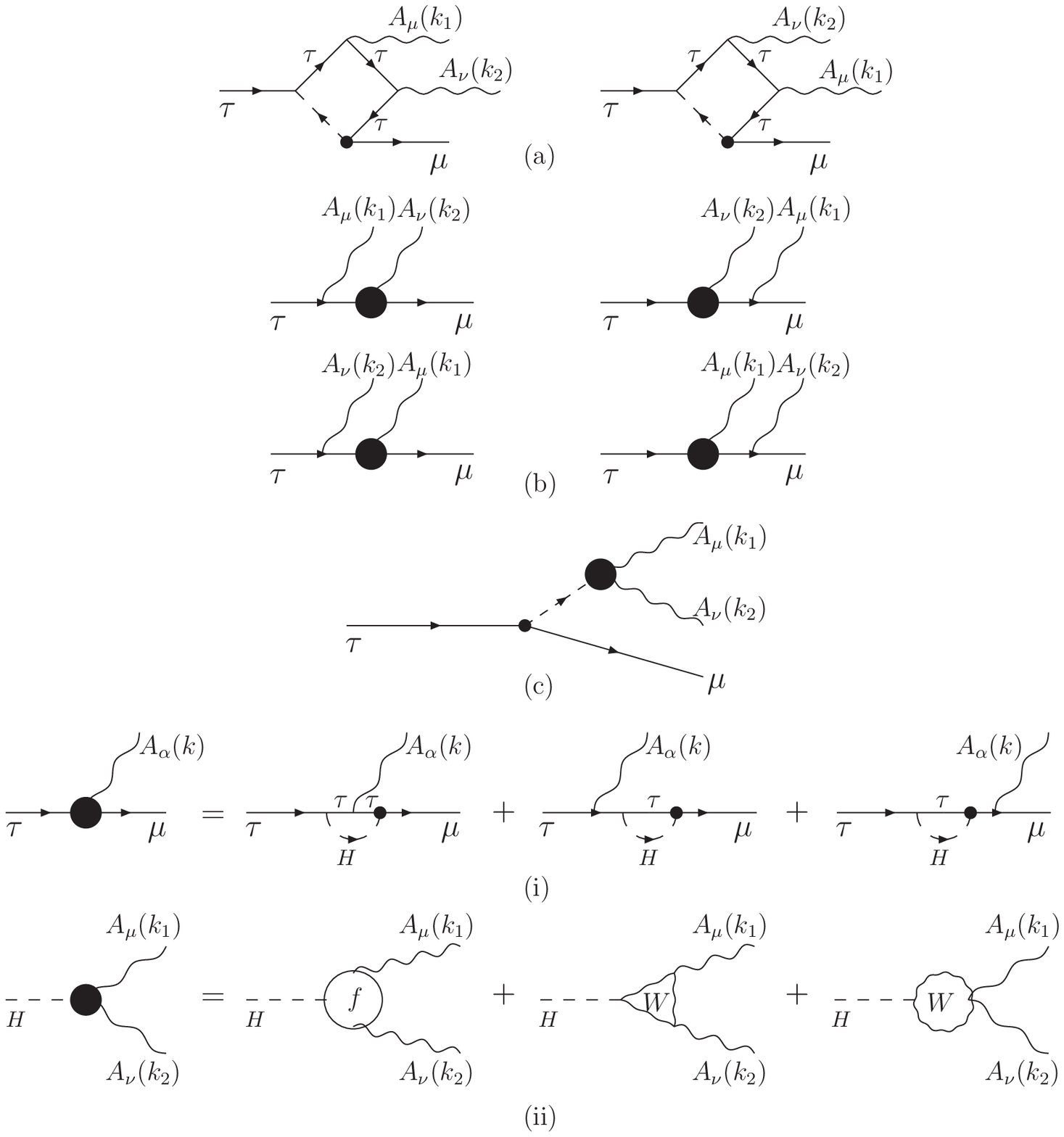}
\caption{\label{G}Feynman diagrams contributing to the $\tau \to
\mu \gamma \gamma$ decay.}
\end{figure}

The two--body $\tau \to \mu \gamma$ decay is induced by the
diagrams shown in Fig.~\ref{G}$(i)$, which lead to a finite and
gauge invariant amplitude of magnetic dipolar type. The
corresponding branching ratio can be written as:
\begin{equation}
Br(\tau \to \mu \gamma)=\frac{{\alpha }^2\,{\Omega }^2}{2048\,{\pi
}^3\,s_W^2}\,\frac{m_\tau^2}{m_W^2}\,\frac{m_\tau}{\,\Gamma_\tau}\,|F|^2,
\end{equation}
where
\begin{equation}
F=\frac{1}{2}+2m^2_\tau
C_0(6)+\frac{m^2_H}{m^2_\tau}\Big(1-2\frac{m^2_\tau}{m^2_H}\Big)\Big(B_0(1)-B_0(2)\Big).
\end{equation}
In this expression, $C_0(6)$ and $B_0(i)$ are Passarino--Veltman
scalar functions, which are defined in Table \ref{PV}. As it is
evident, the loop amplitude $F$ is free of ultraviolet
divergences. This LFV decay has already considered~\cite{KP} in
the context of the Two Higgs doublet model.

As to the $\tau \to \mu \gamma \gamma$ transition, it receives
contributions from box diagrams and reducible graphs characterized
by the off--shell $\tau^*\tau \gamma$ and $H^*\gamma \gamma$
couplings, as shown in Fig.~\ref{G}. Each set of diagrams shown in
this figure leads to a finite and gauge invariant result by
itself. We present exact results for the Box and $H^*\gamma
\gamma$ contributions. For the contribution associated with the
$\tau^*\tau \gamma$ coupling, we will follow the method used in
Ref.~\cite{Yao} to calculate the $b\to s\gamma \gamma$ decay,
which consists in assuming that the $\tau \to \mu \gamma \gamma$
transition is induced by the $\tau \to \mu \gamma$ decay followed
by a bremsstrahlung. Once calculated the loop integrals, one can
write down the corresponding amplitude as follows:
\begin{align}
\mathcal{M}=\frac{\alpha\,g\,\Omega_{\mu\tau}}{8\,\pi\, m_W}\,
\overline{u}(p_1)\,\Big(\Gamma^{\mu
\nu}_{\mathrm{Box}}+\Gamma^{\mu \nu}_{\tau^*\tau
\gamma}+\Gamma^{\mu \nu}_{H^*\gamma
\gamma}\Big)\,u(p_2)\epsilon^*_\mu
(k_1,\lambda_1)\epsilon^*_\nu(k_2,\lambda_2)\label{Amplitud},
\end{align}
where
\begin{equation}
\Gamma^{\mu\nu}_{\mathrm{Box}}=\sum\limits^{16}_{i=1}F_i\,T_i^{\mu\nu}.
\end{equation}
Here, the $T_i^{\mu\nu}$ Lorentz tensors are gauge structures
given by:
\begin{align}
T_1^{\mu\nu}=&\,\frac{g^{\mu\nu}\,k_1\cdot
k_2-k_1^\nu\,k_2^\mu}{k_1\cdot k_2},\ \
T_2^{\mu\nu}=\,\frac{(p_1^\mu\,k_1\cdot k_2-k_2^\mu\,k_1\cdot p_1)(p_1^\nu\,k_1\cdot k_2-k_1^\nu\,k_2\cdot p_1)}{(m_\tau\,\,k_1\cdot k_2)^2},\nonumber\\
T_3^{\mu\nu}=&\,\frac{\pFMSlash{k_1}\,(k_1^\nu\,k_2^\mu-g^{\mu\nu}\,k_1\cdot
k_2)}{m_\tau\,\,k_1\cdot k_2},\ \
T_4^{\mu\nu}=\,\frac{\pFMSlash{k_2}\,(g^{\mu\nu}\,k_1\cdot k_2-k_1^\nu\,k_2^\mu)}{m_\tau\,\,k_1\cdot k_2},\nonumber\\
T_5^{\mu\nu}=&\,\frac{(k_1^\nu\,\pFMSlash{k_2}-\gamma^\nu\,k_1\cdot
k_2) (p_1^\mu\,k1\cdot k_2-k_2^\mu\,k_1\cdot
p_1)}{m_\tau\,(k_1\cdot k_2)^2},\ \
T_6^{\mu\nu}=\,\frac{(k_2^\mu\,\pFMSlash{k_1}-\gamma^\mu\,k_1\cdot
k_2)(p_1^\nu\,k_1\cdot k_2-k_1^\nu\,k_2\cdot
p_1)}{m_\tau^3\,\,k_1\cdot k_2},\nonumber \\
T_7^{\mu\nu}=&\,\frac{\pFMSlash{k_2}\,(p_1^\mu\,k_1\cdot
k_2-k_2^\mu\,k_1\cdot p_1)(p_1^\nu\,k_1\cdot k_2-k_1\nu\,k_2\cdot
p_1)}{m_\tau^3\,\,(k_1\cdot k_2)^2},\ \
T_8^{\mu\nu}=\,\frac{(p_1^\mu\,\pFMSlash{k_1}-\gamma^\mu\,k_1\cdot k_1)(p_1^\nu\,k_1\cdot k_2-k_1^\nu\,k_2\cdot p_1)}{m_\tau^3\,\,k_1\cdot k_2},\nonumber\\
T_9^{\mu\nu}=&\,\frac{\gamma^\mu\,\pFMSlash{k_2}\,k_1^\nu
+\pFMSlash{k_1}\,\gamma^\nu\,k_2^\mu-\pFMSlash{k_1}\,\pFMSlash{k_2}\,g^{\mu\nu}
-\gamma^\mu\,\gamma^\nu\,k_1\cdot k2}{m_\tau^2},\ \
T_{10}^{\mu\nu}=\,\frac{\pFMSlash{k_2}\,\gamma^\nu\,(p_1^\mu\,k_1\cdot k_2-k_2^\mu\,k_1\cdot p_1)}{m_\tau^2\,\,k_1\cdot k_2},\nonumber\\
T_{11}^{\mu\nu}=&\,\frac{\pFMSlash{k_1}\,\gamma^\mu\,(p_1^\nu\,k_1\cdot
k_2-k_1^\nu\,k_2\cdot p_1)}{m_\tau^2\,\,k_1\cdot k_2},\ \
T_{12}^{\mu\nu}=\,\frac{\gamma^\mu\,\pFMSlash{k_2}\,\gamma^\nu\,k_1\cdot
k_2-\pFMSlash{k_1}\,\pFMSlash{k_2}\,\gamma^\nu\,k_2^\mu}{m_\tau\,\,k_1\cdot
k_2},\nonumber \\
T_{13}^{\mu\nu}=&\,\frac{\pFMSlash{k_1}\,\gamma^\mu\,\pFMSlash{k_2}\,
k_1^\nu-\pFMSlash{k_1}\,\gamma^\mu\,\gamma^\nu\,k_1\cdot
k_2}{m_\tau^3},\ \
T_{14}^{\mu\nu}=\,\frac{\,\pFMSlash{k_1}\,\pFMSlash{k_2}\,\gamma^\nu\,
(p_1^\mu\,k_1\cdot k_2-k_2^\mu\,k_1\cdot
p_1)}{m_\tau^3\,\,k_1\cdot k_2},\nonumber \\
T_{15}^{\mu\nu}=&\,\frac{\pFMSlash{k_1}\,\gamma^\mu\,\pFMSlash{k_2}\,
p_1^\nu-\pFMSlash{k_1}\,\gamma^\mu\,\gamma^\nu\,k_2\cdot
p_1}{m_\tau^3},\ \
T_{16}^{\mu\nu}=\,\frac{\pFMSlash{k_1}\,\gamma^\mu\,\pFMSlash{k_2}\,
\gamma^\nu}{m_\tau^2}.\label{estructuras}
\end{align}
Notice that the Ward identities $k_{1\mu}\,T^{\mu
\nu}_i=k_{2\nu}\,T^{\mu \nu}_i=0$ are fulfilled, as required by
electromagnetic gauge invariance. On the other hand, the
contributions arising from the reducible sets of diagrams are
given by:
\begin{align}
\Gamma_{\mathrm{\tau^*\tau \gamma}}^{\mu\nu}=&\,F_{17}
\Bigg[\left(\frac{p_1^\nu}{p_1\cdot k_2}-\frac{p_2^\nu}{p_2\cdot
k_2}\right)
\,\sigma^{\mu\alpha}\,k_{1\alpha}+\left(\frac{p_1^\mu}{p_1\cdot
k_1}-\frac{p_2^\mu}{p_2\cdot k_1}\right)
\,\sigma^{\nu\beta}\,k_{2\beta}\nonumber\\
&-\frac{i}{2}\,\Bigg(\left(\frac{1}{p_1\cdot
k_2}-\frac{1}{p_2\cdot
k_1}\right)\,\sigma^{\nu\alpha}\,\sigma^{\mu\beta}\,k_{2\alpha}\,
k_{1\beta}+\left(\frac{1}{p_1\cdot k_1}-\frac{1}{p_2\cdot
k_2}\right)\,\sigma^{\mu\alpha}\,\sigma^{\nu\beta}\,k_{1\alpha}\,
k_{2\beta}\Bigg)\Bigg],\label{estructura2}
\end{align}
\begin{align}
\Gamma_{\mathrm{H^*\gamma
\gamma}}^{\mu\nu}=\,&F_{18}\,\frac{k_2^\mu\,k_1^\nu-k_1\cdot
k_2\,g^{\mu\nu}}{2\,k_1\cdot
k_2-m_H^2+i\,m_H\,\Gamma_H}.\label{estructura3}
\end{align}
It is easy to see that the Ward identities
$k_{1\mu}\,\Gamma_{\mathrm{\tau^*\tau
\gamma}}^{\mu\nu}=k_{2\nu}\,\Gamma_{\mathrm{\tau^*\tau
\gamma}}^{\mu\nu}=0$ and $k_{1\mu}\,\Gamma_{\mathrm{H^*\gamma
\gamma}}^{\mu\nu}=k_{2\nu}\,\Gamma_{\mathrm{H^*\gamma
\gamma}}^{\mu\nu}=0$ are also satisfied. In the above expressions,
the $F_i$ functions depend on Passarino--Veltman form factors,
which are given in the appendix. As already mentioned, each set of
diagrams shown in Figs. \ref{G}$(a)$, \ref{G}$(b)$, and
\ref{G}$(c)$, leads to a gauge invariant and finite result, as it
can be seen from the expressions given above and those presented
in the appendix. We used the Passarino-Veltman reduction scheme \cite{PASSARINO}
to calculate the amplitudes for each set of diagrams.

Using the above results, the branching ratio for the $\tau \to \mu
\gamma \gamma$ decay can be written as:
\begin{equation}
Br(\tau \to \mu \gamma \gamma)=\frac{\alpha^3\,\Omega^2_{\tau
\mu}}{32\,(4\pi)^4\,s^2_W}\,\frac{m_\tau^2}{m_W^2}\,\frac{m_\tau}{\Gamma_\tau}\,
\int\limits^1_\frac{2m_\mu}{m_\tau}dx\int\limits^{1}_{1-x}dy\,f(x,y),
\end{equation}
where $f(x,y)=(32\pi s^2_W/\alpha^3\Omega^2_{\tau
\mu})(m_W/m_\tau)^2|\overline{\cal M}|^2$. In addition, the
scaled variables $x=2k_1\cdot k_2/m^2_\tau$ and
$y=2p_2\cdot k_2/m^2_\tau$ were introduced \cite{JPG-TAVARES}.

\begin{figure}
\centering
\includegraphics[width=2.5in]{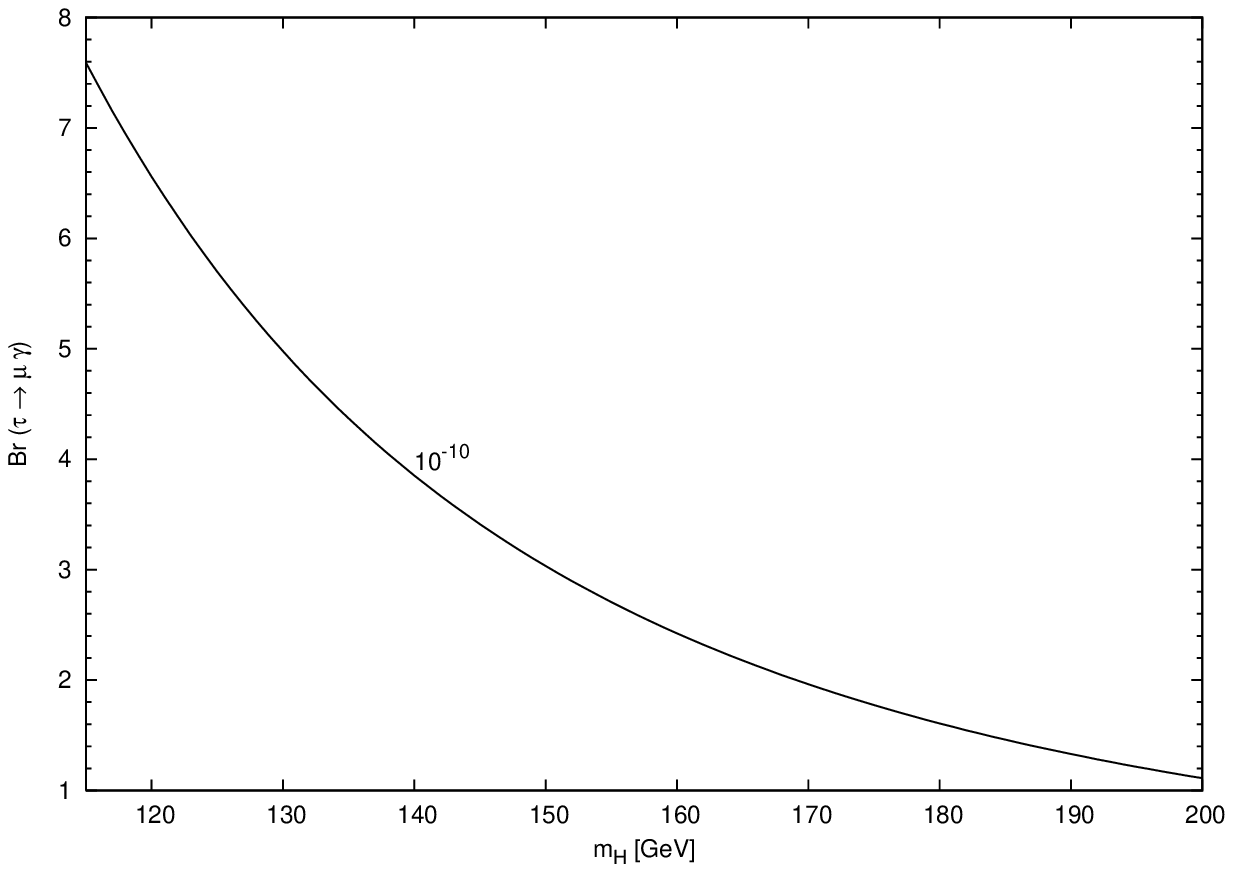}
\includegraphics[width=2.5in]{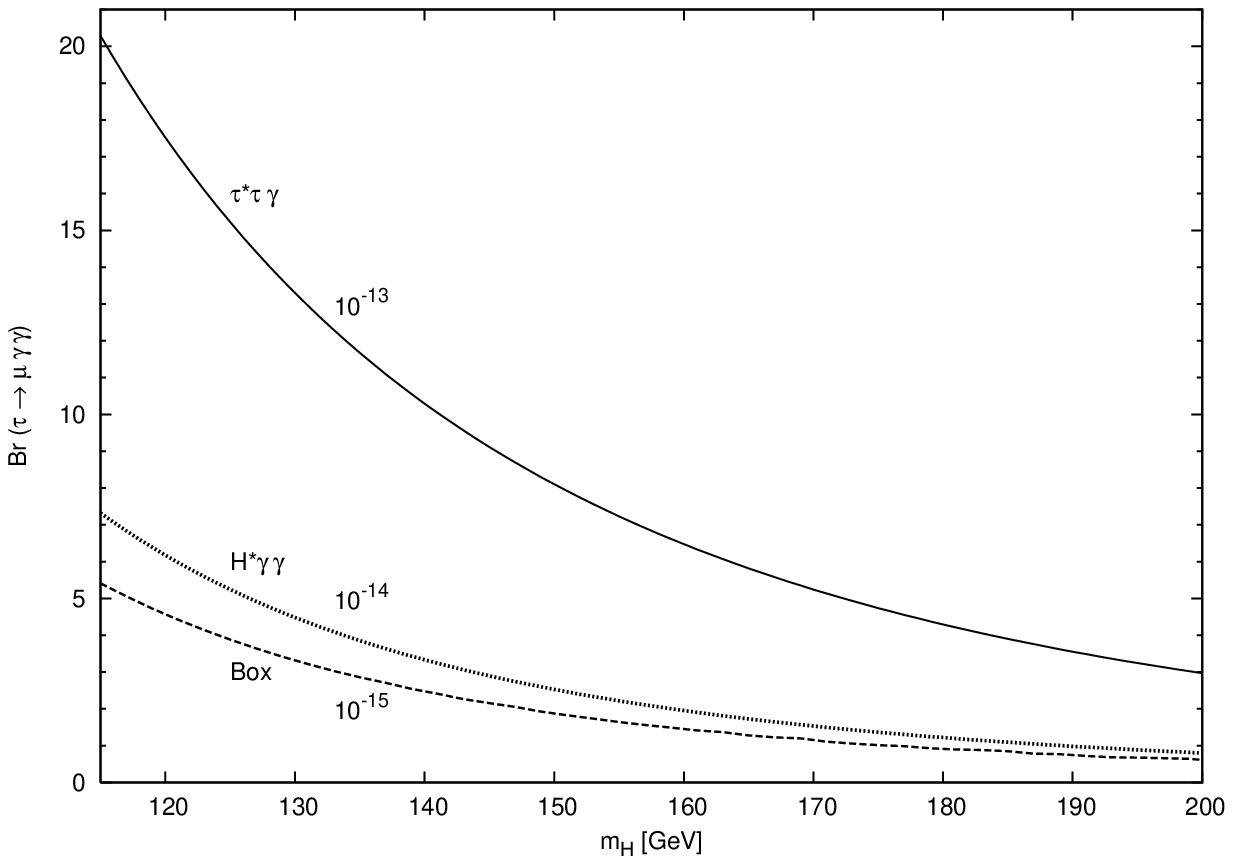}
\caption{\label{23b}Branching ratio for the $\tau \to \mu \gamma$
and $\tau \to \mu \gamma \gamma$ decays as a function of the Higgs
mass.}
\end{figure}

In order to make predictions, we need to assume some value for the
$\Omega_{\tau \mu}$ parameter. Low--energy data can be used to
constraint it. It was shown in Ref.~\cite{FPT} that the best
bound, namely $|\Omega_{\tau \mu}|^2<2\times 10^{-4}$, arises from
the experimental value on the muon anomalous magnetic
moment~\cite{PDG}\footnote{A similar bound is obtained from the
experimental limit on the LFV $\tau \to \bar{\mu}\mu \mu$
decay~\cite{FPT}.}. In Fig.~\ref{23b} the branching ratios for the
$\tau \to \mu \gamma$ and $\tau \to \mu \gamma \gamma$ decays are
shown as a function of the Higgs mass. In the case of the
three--body $\tau \to \mu \gamma \gamma$ decay, we have displayed
separately the contributions coming from box diagrams and
reducible graphs characterized by the $\tau^*\tau \gamma$ and
$H^*\gamma \gamma$ couplings. From this figure, it can be
appreciated that the contribution induced by the $\tau^*\tau
\gamma$ coupling is larger than those generated by the $H^*\gamma
\gamma$ coupling and Box diagrams in approximately one and three
orders of magnitude, respectively. On the other hand, it can be
seen that these branching ratios varies in less of one order of
magnitude for $m_H$ ranging from 115 to 200 GeV. The
largest values for the branching ratios, $Br(\tau \to \mu \gamma)<2.5\times 10^{-10}$ and
$Br(\tau \to \mu \gamma \gamma)<2.3\times 10^{-12}$, are reached
for the lowest value of the Higgs mass allowed by the LEP bound~\cite{ALEPH}, namely, $m_H=114.4$ GeV.

It is interesting to compare our predictions for $Br(\tau \to \mu
\gamma)$ and $Br(\tau \to \mu \gamma \gamma)$ with current
experimental limits on this class of transitions. The Review of
Particle Physics~\cite{PDG} reports the limit $Br(\tau \to \mu
\gamma)<6.8\times 10^{-8}$, which is two orders of magnitude less
stringent that our bound. As to three--body decays, only there is
experimental information on the muon decay $\mu \to e\gamma
\gamma$, whose branching ratios cannot be larger than $7.2\times
10^{-11}$, which is almost two orders of magnitude lower than our
bound for the analogous tau decay.

In conclusion, in this paper the electromagnetic transitions $\tau
\to \mu \gamma$ and $\tau \to \mu \gamma \gamma$ induced at
one--level by a LFV $H\tau \mu$ vertex were studied in a
model--independent way using the effective Lagrangian approach. Analytical expressions for the contributions induced by the box diagrams were presented. Upper bounds for the corresponding branching ratios were derived
using low--energy data. These bounds are more stringent than
current experimental limits on this class of transitions. Our
results suggest that if LFV electromagnetic transitions of the tau
lepton are detected, they would be generated by sources different
from extended Yukawa sectors.

\acknowledgments{We acknowledge financial support from CONACYT and
SNI (M\' exico).}

\appendix
\section{The Passarino--Veltman Form factors}
Using the notation given in Ref.~\cite{FC} for the
Passarino--Veltman form factors, the $F_i$ functions can be
written as:
\begin{align}
F_1=\,&-2\,m_\tau^2\,\Big(C_0(4) + m_\tau^2\,x\,D_0(1)-
2\,\big(D_{00}(1) + D_{00}(2) \big)\Big),\ \ \
F_2=\,4\,m_\tau^4\,
\Big(D_{22}(1) + D_{22}(2)\Big),\nonumber\\
F_3=\,&-2\,m_\tau^2\,\Big(C_0(5) + C_2(1)-
m_\tau^2\,x\,\left(D_0(1) + D_1(1) + D_2(1) + D_3(1) \right) - 2\,\big(D_{00}(2)\nonumber\\
&- D_{001}(1) + D_{001}(2) - D_{002}(1) - D_{003}(1) + D_{003}(2)\big)\Big),\nonumber\\
F_4=\,&-2\,m_\tau^2\,\Big(C_2(1) + 2\,\big(D_{00}(1) +
D_{001}(1) - D_{001}(2) + D_{002}(1)\big)\Big),\nonumber \\
F_5=\,&-2\,m_\tau^2\,\Big( C_0(5) + C_1(3) + C_2(3) +
2\,\big(D_{002}(1) +  D_{002}(2) \big)  \Big), \nonumber \\
F_6=\,&-4\,m_\tau^4\,\Big(D_{12}(1) + D_{22}(1) + D_{112}(1)
+ D_{112}(2) + 2\,D_{122}(1) + D_{123}(1) + D_{123}(2)+
D_{222}(1) + D_{223}(1)\Big),\nonumber\\
F_7=\,&-4\,m_\tau^4\,\Big(D_{22}(1) + D_{122}(1) -
D_{122}(2) +
D_{222}(1)\Big),\nonumber\\
F_8=\,&-4\,m_\tau^4\,\Big(D_{22}(1) + D_{122}(1) - D_{122}(2)
+ D_{222}(1) + D_{223}(1) -
D_{223}(2)\Big),\nonumber\\
F_9=\,&-2\,m_\tau^4\,D_0(1),\ \
F_{10}=\,2\,m_\tau^4\,\left(D_{2}(1) + D_2(2)\right),\ \
F_{11}=\,2\,m_\tau^4\,\left(D_2(1) +
D_2(2)\right),\nonumber \\
F_{12}=\,\,&m_\tau^2\,\Big(C_0(3) - C_0(4) + \big(m_H^2 -
2\,m_\tau^2\big)\,D_0(2) + C_1(2) + m_\tau^2\,\left( 2 + x \right)
\,D_1(2) - C_2(1)\nonumber\\ &+ m_\tau^2\,\left(x + y
-2\right)\,D_2(2) - m_\tau^2\,\left( x + y -2\right)\,D_3(2) -
2\,D_{00}(1) + 2\,D_{00}(2)\nonumber\\ &+ m_\tau^2\,x\,D_{11}(2) +
m_\tau^2\,\left(x -2  \right)\,D_{12}(2) +
m_\tau^2\,x\,D_{13}(2) + m_\tau^2\,\left( x + y  -2\right)\,D_{23}(2)\Big),\nonumber\\
F_{13}=\,&-2\,m_\tau^4\,\Big(D_1(1) + D_1(2) + D_2(1) +
D_{11}(1) + D_{11}(2) + 2\,D_{12}(1) +
D_{13}(1)\nonumber\\ &+ D_{13}(2) + D_{22}(1) + D_{23}(1)\Big),\nonumber\\
F_{14}=\,&-2\,m_\tau^4\,\Big(D_2(1) + D_{12}(1) - D_{12}(2) +
D_{22}(1) + D_{23}(1) - D_{23}(2)\Big),\nonumber \\
F_{15}=\,&-2\,m_\tau^4\,\Big(D_2(1) + D_{12}(1) - D_{12}(2) +
D_{22}(1)\Big),\ \
F_{16}=\,m^4_\tau\,\left(D_0(1)+D_0(2) \right),\nonumber\\
F_{17}=\,&\frac{1}{2\,m_\tau^2}\Big(m_\tau^2 + 2\,\left(m_H^2 -
2\,m_\tau^2 \right) \,
\left(B_0(1) - B_0(2)\right) + 4\,m_\tau^4\,C_0(6)\Big),\nonumber\\
F_{18}=\,&\frac{8\,m_W^2}{m_\tau^2\,x}\left(3 +
\frac{m_\tau^2\,x}{2\,m_W^2} + 6\,m_W^2\left(1 -
\frac{m_\tau^2\,x}{2\,m_W^2}\right)\,C_0(1)\right)
-Q_t^2\,{N_{c_{t}}}\,\frac{8\,m_t^2}{m_\tau^2\,x} \left(2 +
(4\,m_t^2 - m_\tau^2\,x)\,C_0(2)\right).
\end{align}

\begin{table}
\begin{center}
\begin{tabular}{@{}ccccccccccc@{}}
\hline $(i)$&1&2&3&4&5&6&7&8&9&10\\ \hline
$B_0(1)$&$0$&$m_H^2$&$m_\tau^2$&$$&$$&$$&$$&$$&$$&$$\\
$B_0(2)$&$m_\tau^2$&$m_H^2$&$m_\tau^2$&$$&$$&$$&$$&$$&$$&$$\\
$C_0(1)$&$0$&$0$&$x\,m_\tau^2$&$m_W^2$&$m_W^2$&$m_W^2$&$$&$$&$$&$$\\
$C_0(2)$&$0$&$0$&$x\,m_\tau^2$&$m_t^2$&$m_t^2$&$m_t^2$&$$&$$&$$&$$\\
$C_0(3)$&$0$&$0$&$x\,m_\tau^2$&$m_\tau^2$&$m_\tau^2$&$m_\tau^2$&$$&$$&$$&$$\\
$C_0(4)$&$0$&$2\,m_\tau^2$&$x\,m_\tau^2$&$m_\tau^2$&$m_H^2$&$m_\tau^2$&$$&$$&$$&$$\\
$C_0(5)$&$0$&$2\,m_\tau^2$&$y\,m_\tau^2$&$m_\tau^2$&$m_\tau^2$&$m_H^2$&$$&$$&$$&$$\\
$C_0(6)$&$m_\tau^2$&$0$&$0$&$m_H^2$&$m_\tau^2$&$m_\tau^2$&$$&$$&$$&$$\\
$C_l,C_{lm}(1)$&$0$&$x\,m_\tau^2$&$2\,m_\tau^2$&$m_H^2$&$m_\tau^2$&$m_\tau^2$&$$&$$&$$&$$\\
$C_l,C_{lm}(2)$&$2\,m_\tau^2$&$x\,m_\tau^2$&$0$&$m_H^2$&$m_\tau^2$&$m_\tau^2$&$$&$$&$$&$$\\
$C_l,C_{lm}(3)$&$y\,m_\tau^2$&$0$&$2\,m_\tau^2$&$m_H^2$&$m_\tau^2$&$m_\tau^2$&$$&$$&$$&$$\\
$D_0,D_{l},D_{lm},D_{lmn}(1)$&$x\,m_\tau^2$&$0$&$y\,m_\tau^2$&$0$&$2\,m_\tau^2$&$0$&$m_\tau^2$&$m_\tau^2$&$m_H^2$&$m_\tau^2$\\
$D_0,D_{l},D_{lm},D_{lmn}(2)$&$x\,m_\tau^2$&$2\,m_\tau^2$&$m_\tau^2(2-x-y)$&$0$&$0$&$0$&$m_\tau^2$&$m_\tau^2$&$m_H^2$&$m_\tau^2$\\
\hline
\end{tabular}
\caption{The arguments for the Passarino--Veltman form factors.}
\label{PV}
\end{center}
\end{table}

\end{document}